 \documentclass[secnumarabic, twocolumn, graphics,floatfix, superscriptaddress,tightenlines, aps, prb]{revtex4-1}
\pdfoutput=1
\usepackage{graphicx}
\usepackage{physics}
\usepackage{color}
\usepackage{amsmath}
\usepackage{fullpage}
\usepackage{hyperref}
\usepackage{float}

\hypersetup{
    colorlinks=true,       
    linkcolor=black, 
    citecolor=blue, 
    filecolor=magenta,      
    urlcolor=cyan           
}

\begin{document}
\title{Transition from spin-orbit to hyperfine dominated spin relaxation in a cold fluid of dipolar excitons}
%
%
\author{Ran Finkelstein}
\altaffiliation{Current Address:{ Department of Physics of Complex Systems, Weizmann Institute of Science, Rehovot 76100, Israel}}
\affiliation{Racah Institute of Physics, The Hebrew University of Jerusalem, Jerusalem 9190401, Israel}

\author{Kobi Cohen}
\affiliation{Racah Institute of Physics, The Hebrew University of Jerusalem, Jerusalem 9190401, Israel}
\author{Benoit Jouault}
\affiliation{Laboratoire Charles Coulomb, UMR 5221 CNRS-Universit\'e de Montpellier, Montpellier, F-34095, France}
\author{Ken West}
\affiliation{Department of Electrical Engineering, Princeton University, Princeton, New Jersey 08544, USA.}
\author{Loren N. Pfeiffer}
\affiliation{Department of Electrical Engineering, Princeton University, Princeton, New Jersey 08544, USA.}
\author{Masha Vladimirova}
\affiliation{Laboratoire Charles Coulomb, UMR 5221 CNRS-Universit\'e de Montpellier, Montpellier, F-34095, France}
\author{Ronen Rapaport}
\affiliation{Racah Institute of Physics, The Hebrew University of Jerusalem, Jerusalem 9190401, Israel}
\affiliation{Applied Physics Department, The Hebrew University of Jerusalem, Jerusalem 9190401, Israel}

\begin{abstract} 
We measure the spin-resolved transport of dipolar excitons in a biased GaAs double quantum well structure. From these measurements we extract both spin lifetime and mobility of the excitons. We find that below a temperature of $4.8$K, there is a sharp increase in the spin lifetime of the excitons, together with a sharp reduction in their mobility. 
%
At $T=1.5$K, where the excitons have the lowest mobility, we observe an anomalous non-monotonous dependence of the exciton spin lifetime on the mobility. 
Below a critical power the spin lifetime increases with increasing mobility and density, while above the critical power the opposite trend is observed.
We interpret this transition as an evidence of the interplay between two different spin dephasing  mechanisms: at low mobility the dephasing is dominated by the hyperfine interaction with the lattice nuclei spins, while at higher mobility the spin-orbit interaction dominates, and a Dyakonov-Perel spin relaxation takes over.
The excitation power and temperature regime where the hyperfine interaction induced spin dephasing is observed correlates with the regime where a dark dipolar quantum liquid was reported recently on a similar sample.   
\end{abstract}
\maketitle

\section{Introduction}

The spin degree of freedom plays a crucial role in electronic, magnetic, optical and transport properties of semiconductor nanostructures, and spin is also one of the most promising candidates for encoding and transporting quantum information in such structures. The state of an electron spin is very sensitive to interactions with its environment. 
Two main interactions are known to affect electron spin:  spin-orbit interaction, that couples the spin of  electron to its motion, and hyperfine interaction that couples electron spin to the spin of the lattice nuclei. 
 They give rise to striking phenomena such as the spin Hall effect \cite{Wunderlich2005}, persistent spin helix \cite{Koralek2009} and self-sustained polarization oscillations \cite{OpticalOrientation}, but on the same time they are the two main sources of spin relaxation and dephasing \cite{KavokinSST,Dyakonov2008a}. Relative importance of these two types of interaction is related to the electron mobility \cite{Dzhioev2002,Chen2007}. 

This fact is particularly interesting for an electron bound in a dipolar, or  indirect exciton (IX), a Coulomb-bound pair made of an electron and a hole, where each constituent is confined in a separate GaAs quantum well separated from the other by a thin (Al, Ga)As barrier. Such IXs form a long living two-dimensional fluid of interacting boson-like particles, with a complex, four-fold internal spin structure. 
Two of the IX spin states can couple to the light (bright states), they have spin projection along the growth axis $S_z=\pm1$, 
while two other spin states ($S_z=\pm2$) are dark.
It was predicted that Bose-Einstein condensation of the IX gas into the dark spin state should occur below a critical temperature \cite{Combescot2007}, and that the condensed phase should be strongly correlated due to the strong dipolar interactions between the particles, similarly to a quantum liquid \cite{Laikhtman2009,Lozovik_SSC}.
Recent experiments on dipolar IXs have reported a condensation to a dark, correlated quantum liquid-like phase \cite{Shilo2013,Stern2014,Cohen2016}, with a strongly quenched IX motion \cite{Stern2014,Cohen2016,Mazuz-Harpaz2017}, as well as reports on a condensate with an extended spatial coherence \cite{High2012a,High2012,Alloing2013a} and a ballistic spin motion \cite{High2013}. 

\begin{figure}
\includegraphics[scale=0.3]{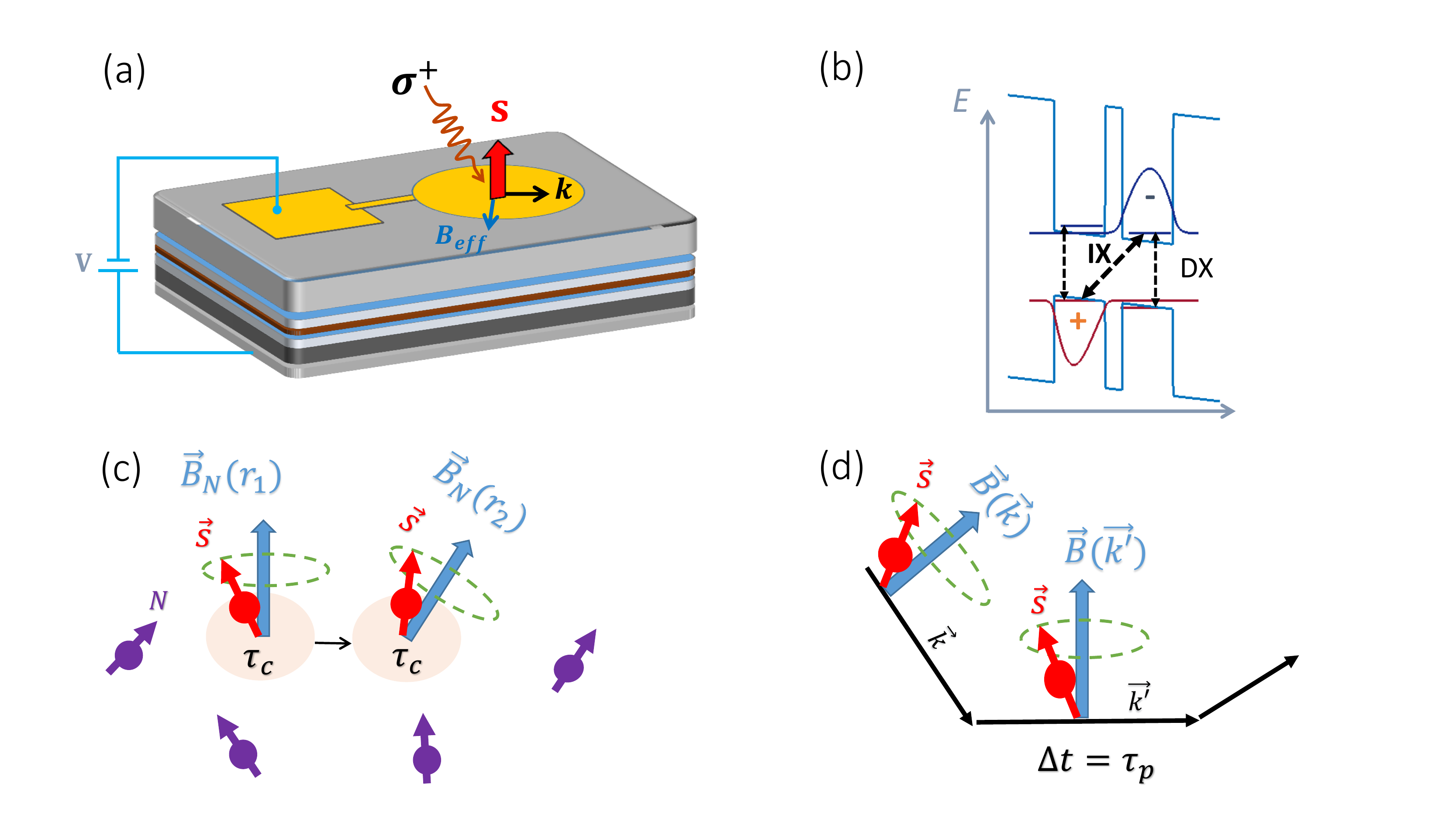}\protect\caption{(a) Sketch of the sample and experimental setup. Coupled quantum wells subjected to an electric field in the z direction are excited with circularly polarized light resonant with the DX energy.(b) Band diagram of the biased GaAs/AlGaAs coupled quantum wells. A resonant excitation of the DX transition is followed by an electron tunneling across the barrier, and the formation of IXs. (c) Hyperfine interaction-induced spin relaxation for localized excitons:  relaxation rate is inversely proportional to the electron spin correlation time $\tau_c$. This mechanism is quenched for mobile excitons; (d) Dyakonov-Perel spin relaxation mechanism for moving excitons via fluctuations of the effective magnetic field induced by spin-orbit interaction. Relaxation rate is 
proportional to the momentum scattering time $\tau_p$ }
\label{fig: intro}
\end{figure}

There are many open questions on the properties, the microscopic nature, and the formation dynamics of the newly observed collective phase. For example, the darkening of the IX fluid is suggested to be attributed to a spontaneous condensation of IX to the low energy dark spin states \cite{Combescot2007,Shilo2013,Cohen2016,Alloing2013a}.  In this context, the spin of the electron inside the IX could in principle be such a complementary, sensitive probe of IX dynamics. 
Indeed, long spin lifetimes (up to tens of nanoseconds) have been reported for IX  \cite{Violante2015,Leonard2009,Kowalik-Seidl2010,Beian2015,Andreakou2015,Andreakou2016}. This is due to dramatic reduction of the electron-hole exchange interaction within IX, 
which is responsible for the fast spin relaxation of traditional, direct excitons (DXs) in semiconductor quantum wells \cite{Vinattieri1993}. Because hole spin relaxation is usually much faster than spin relaxation of an electron, IX spin lifetime is determined by the constituent electron spin lifetime.
Thus, for IX spin relaxation we can expect a phenomenology similar to that of a two-dimensional electron gas.  
 When IXs are localized ({\it e. g.} as observed in low mobility correlated liquid of IXs, or in the case of strong disorder), the spin dephasing could be dominated by the hyperfine interaction, via the fluctuations of the Overhauser field $B_N$ \cite{OpticalOrientation,KavokinSST,Dyakonov2008a}. This mechanism is schematically represented in Fig.~\ref{fig: intro}~(d): electron spin relaxation rate is determined by the electron correlation time $\tau_c$ (during which the Overhauser field experienced by an electron can be considered as constant) and the squared average value of $B_N$.  In contrast, in a regime where IXs are highly mobile, fluctuations of the Overhauser field are averaged out. In this case electron spin dephasing is mainly due to spin-orbit interaction, via Dyakonov-Perel (DP) mechanism (Fig.~\ref{fig: intro}~(c)) \cite{Dyakonov1971}. Here the relevant time is the momentum scattering time $\tau_p$.

 A natural way to probe these two regimes is to study the correlation between IX spin lifetime and mobility, by changing the lattice temperature and the exciton density.
To get access to both spin lifetime and mobility, spatial extension of the IX cloud upon point-like excitation, and its polarization degree should be analyzed in the same experiment. Previous works have reported a significant decrease of the IX spin lifetime with increasing IX diffusivity, a behavior consistent with DP relaxation mechanism of IX \cite{Leonard2009}. However, the role of the Overhauser field in the exciton spin dynamics has not been evidenced so far. Indeed,  when  hyperfine mechanism of spin relaxation dominates, one should expect an increase of the IX spin lifetime with increasing IX diffusivity and density.



In this work we report, for the first time in any excitonic system to our knowledge, a temperature and density regime where a transition from a decreasing spin lifetime to an increasing spin lifetime with increasing mobility takes place.
This suggests that hyperfine interaction does play a role in the IX spin relaxation when a transition from high to low mobility state of the IX fluid occurs.  Remarkably, the transition temperature between the two regimes is found to closely match the previously reported critical temperature for the condensation of an IX liquid.

The paper is organized as follows: The sample and experimental setup are presented in Section \ref{setup}. It is followed by the description of 
experimental results (Section \ref{results}), as well as the drift-diffusion model and its assumptions  (Section \ref{model}). Section \ref{discussion} contains the discussion of the observed power and temperature dependences 
and their interpretation in terms of interplay between spin-orbit and hyperfine interaction. The conclusions are given in Section \ref{conclusions}.

\begin{figure*}[h!]
\includegraphics[scale=0.38]{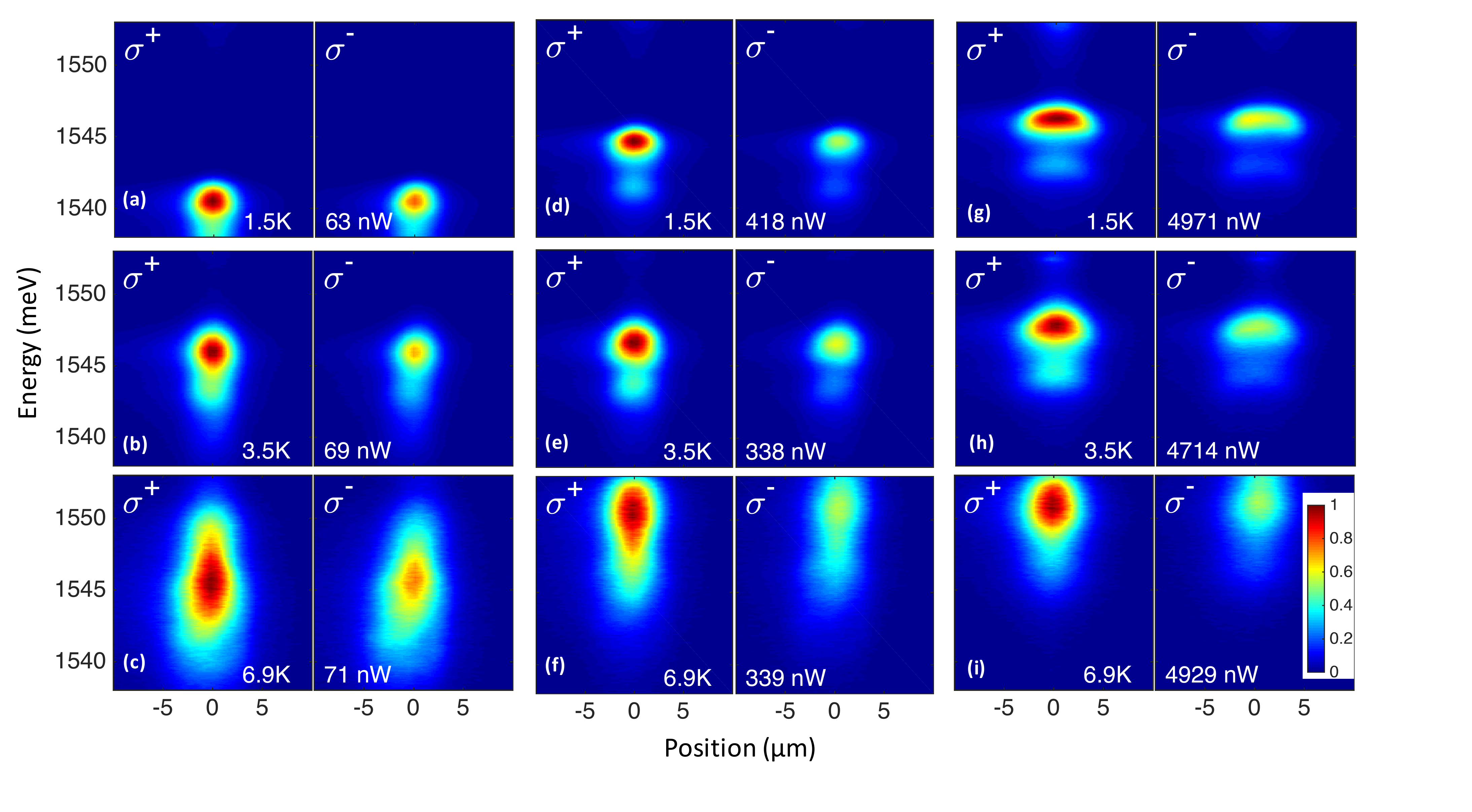}\protect\caption{
Color-encoded spatially resolved PL spectra of IX in the two orthogonal circular polarizations. The measurements at three different excitation powers and three temperatures are shown. Each data set is normalized to its maximum in $\sigma^+$ polarization.}
\label{fig:raw data}
\end{figure*}

 \begin{figure*}[h!]
\centering{}\includegraphics[scale=0.52]{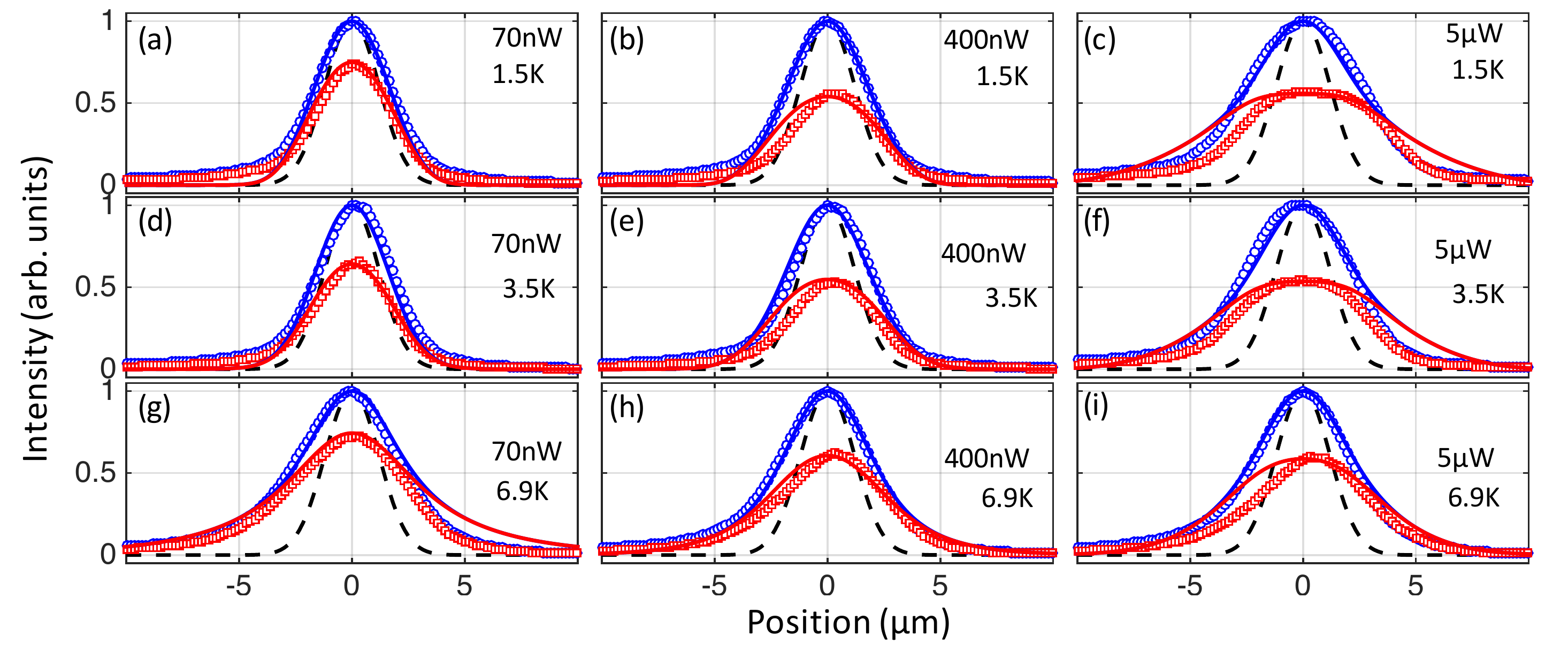}\caption{Spatial distribution of IX emission intensity at several temperatures and excitation powers. $\sigma^+$ (blue circles) and $\sigma^-$ (red squares) intensity profiles are compared to the excitation spot profile (black dashed line). The solid lines are a fit to the model presented in Section \ref{model}.}
\label{figprofiles}
\end{figure*}

\section{Sample And Experimental Setup}
\label{setup}
The sample that we use in this work has already been studied in Refs. \onlinecite{Shilo2013,Cohen2016}. It contains a double quantum well (DQW) structure with two $12$~nm-wide GaAs quantum wells separated by a $4$~nm-wide Al$_{0.45}$Ga$_{0.55}$As barrier. An electric field is applied along the growth axis between a semi-transparent $10$~nm-thick Ti top circular gate with a diameter of $300$~$\mu$m and n$^+$-doped GaAs substrate, as depicted in Fig.~\ref{fig: intro}(a). Fig.~\ref{fig: intro}(b) shows the band diagram  of the biased DQW structure with direct exciton (DX) and IX transitions. A tunable Titanium-Sapphire laser is focused through a $\times20$ objective (numerical aperture NA=0.42) to a beam of ~ $3$~$\mu m$ diameter at the sample surface. The sample is mounted inside an immersion He$^4$ optical cryostat.
The photoluminescence (PL) 
is collected through the same objective, and split into two beams with orthogonal polarizations ($\sigma^+$ and $\sigma^-$), simultaneously imaged onto a spectrometer with magnification of $40$. The spectrometer includes a CCD with a pixel size of 26~$\mu$m, such that the spatial imaging resolution is $0.65$~$\mu$m.
We use photoluminescence excitation (PLE) experiments to determine the  energy of the DX state in our sample,  $E_{DX}=1.558$~eV (see Appendix).
To optimize the spin pumping efficiency, we tune a $\sigma^+$-polarized excitation exactly to that energy. The excitation power is varied, resulting in variation of the resulting IX densities.  The applied bias is fixed at $V_g=3$~V, corresponding to the IX emission energy $E=1.5408$~eV at the lowest measured temperature and excitation power. 
To determine the radiative lifetime  $\tau_r$ of the IXs, we directly measured $\tau_r^0$ for a specific DX-IX energy separation $\Delta^0_\mathrm{DX-IX}=26$~meV, using a time-resolved PL experiment. A radiative lifetime of $\tau_r^0=460$~ns was found. To extract $\tau_r$ for all other experiments, we use the quadratic analytic model which was recently shown to be a good approximation for the dependence of $\tau_r$ on the DX-IX energy splitting $\Delta_\mathrm{DX-IX}$: \cite{Shilo2013,Mazuz-Harpaz2017a}
 \begin{equation}
 \tau_r=\tau_r^0 \left( \frac{ \Delta_\mathrm{DX-IX}}{\Delta^0_\mathrm{DX-IX}}\right)^2.
 \label{taur}
 \end{equation}
%
\section{Results}
\label{results}
Fig. \ref{fig:raw data} shows typical color maps of a spatial cross-section of PL spectra as a function of the distance from the excitation spot for two circular polarizations. These measurements are taken at three different excitation powers and three temperatures. For each 
pair of measurements $\sigma^+/\sigma^-$ the intensity is normalized to its maximum in $\sigma^+$ polarization.
Several observations could be made from these measurements: (i) The IX emission energy 
increases with increasing excitation power for each temperature. This is a signature of the increasing IX density. This energy shift results in the reduction of the 
IX lifetime $\tau_r$ (see Eq. \ref{taur} and Fig. \ref{figP}~(a));  
(ii) the IX emission is partially polarized, which means that  the transfer between DX and IX states partially conserves the polarization, in agreement with previous reports \cite{Poggio2004,Leonard2009,Violante2015,Kowalik-Seidl2010,Beian2015}; (iii) there are variations in the spatial extent, polarization and shape of the IX cloud for different excitation powers. This is a signature of the dipole-dipole interactions between IXs. There is a non-trivial dependence of the diffusion coefficient on the IX density and temperature, due to an interplay between drift and diffusion currents of IX \cite{Rapaport2006}.
To get a closer look on the shape of the IX emission intensity profile and its polarization 
we plot in Fig. \ref{figprofiles} the spectrally integrated intensity of the IX
emission in $\sigma^+$ (blue symbols) and  $\sigma^-$ (red symbols) polarizations for the same set of measurements as those shown in Fig.~\ref{fig:raw data}. The excitation spot size is shown by a black dashed line. 
To extract the IX spin lifetime and diffusivity from these data sets, we fit them to the drift-diffusion 
model presented in the next Section.

%
\section{Model}
\label{model}
 We consider the simplest model that accounts for radial drift, diffusion, 
and spin relaxation of IXs created by circularly polarized pump beam with a $R_0$-radius Gaussian spot. The same model was used in Ref. \onlinecite{Leonard2009}.

Let $n=[n_{+2}  \: n_{+1} \: n_{-1}  \: n_{-2}]$ be the vector of IX densities in each of four IX spin states, $D$ is the IX diffusion coefficient. IX mobility $\mu$ is related to the diffusion coefficient via  Einstein relation (linear transport regime \cite{Rapaport2005}), $\mu=D/k_B T$, where $k_B$ is the Boltzmann constant and $T$ is the lattice temperature. The potential  $u_0$ responsible for the drift of IXs out from the excitation spot is due to the dipole-dipole interaction between IXs \cite{Ivanov2002,Rapaport2005,Rapaport2006,Rapaport2007}.
In the simplest case $u_0$ is  given by the plain capacitor formula $u_0=4 \pi^2 d / \epsilon$, where $d$ is the distance between the quantum well centers, and $\epsilon$ is the dielectric constant \cite{Ivanov2002,Laikhtman2009,Schindler2008,Remeika2015}. For the samples studied in this work, this yields $u_0=1.6\times 10^{10}$~meV~cm$^2$. The exciton generation rate under $\sigma^{+1}$ excitation in the sample plane can be written as  
$\Lambda_{+1}= 2  (N_p / R_0^2)  \exp(-r^2/R_0^2)$, where $r$ is the in-plane distance from the excitation spot center, and $N_p$ is the number of excitons generated per second. 

 The drift-diffusion equation including spin dynamics of the four exciton states reads
 \begin{equation}
 \frac{dn}{dt}=\nabla(D \nabla n)+ \mu n \nabla (u_0 (n_{b}+n_{d}))+w n +\Lambda,
\end{equation}
where $n_b+n_d=(n_{+2}+n_{+1}+n_{-1} +n_{-2})$ is the total (bright and dark) exciton density, and $\Lambda=[0 \:\Lambda_{+1}  \:0\: 0$] is the generation rate vector. The exciton recombination and spin relaxation (including individual electron and hole spin flips, as well as the exciton spin flips) is represented by the matrix $w$\cite{Maialle1993}:
 \begin{widetext}
\begin{equation}
 w= \left(
 \begin{array}{cccc}
 -(w_e^+ + w_h^+) & w_e^- & w_h^-& 0 \\
 w_e^+ & -(\frac{1}{\tau_r} +w_e^- + w_h^-+ w_{ex}) & w_{ex} &w_h^+ \\
 w_h^+& w_{ex}&  -(\frac{1}{\tau_r} +w_e^- + w_h^-+ w_{ex})  & w_e^+ \\
 0 & w_h^- &  w_e^-& -(w_e^+ + w_h^+) %
\end{array}%
 \right) \label{eq:w}
 \end{equation}
  \end{widetext}
Here $w_{ex}=1/2\tau_{ex}$ is exciton spin flip rate between $S_z=\pm1$ states, $w^\pm_{e(h)}=\tau_{e(h)}^{-1}/(1+\exp(\pm \Delta/k_BT))$ is electron (hole) spin flip rate, and $\Delta$ is the splitting between $S_z=\pm1$ (bright) and $S_z=\pm2$ (dark) states. As $\Delta \ll k_BT$ in the range of temperatures of interest we assume that electron(hole) spin flip rate is simply given by $w_{e(h)}=\tau_{e(h)}^{-1}/2$. 

Assuming (i) fast hole relaxation $\tau_h \ll \tau_e,\tau_{ex}$  and (ii) very slow exciton relaxation $\tau_{ex} \gg \tau_e$, Eq.~\ref{eq:w} can be simplified. For the bright exciton densities we get:
 \begin{eqnarray}
 \begin{split}
  2\frac{dn_{+1}}{dt}=2\nabla(D \nabla n_{+1}+ 2\mu n_{+1} \nabla (u_0 n_{b})) \\ 
-\frac{1}{\tau_r}n_{+1} - \frac{1}{\tau_e}(n_{+1}-n_{-1} )+\Lambda,
\end{split}
 \label{eq:npm}
 \end{eqnarray} 
 \begin{eqnarray}
\begin{split}
 2\frac{dn_{-1}}{dt}=2\nabla(D \nabla n_{-1}+ 2\mu n_{-1} \nabla (u_0 n_{b}))  \\ 
 -\frac{1}{\tau_r}n_{-1}- \frac{1}{\tau_e}(n_{-1}-n_{+1} ).
 \end{split}
  \label{eq:npm2}
 \end{eqnarray} 
Note that these equations do explicitly account for the dark excitons dynamics in the structure, but they assume that bright-dark conversion is infinitely fast, so that $n_{+2}=n_{-1}$ and $n_{-2}=n_{+1}$. One can see from Eqs.~\ref{eq:npm}, \ref{eq:npm2} that within the assumptions made here, the polarization lifetime (the time corresponding to the transition from one bright IX state to another) is simply equal to the electron spin-flip time $\tau_s=\tau_e$.  In the following we will refer to this time $\tau_s$ as the IX spin lifetime.

In order to find the steady state solutions for the exciton density, we numerically solve this set of equations. The full width at half maximum (FWHM) of the Gaussian excitation spot radius is extracted from the fit to the Gaussian function of the laser spot profile (dashed lines in Fig.~\ref{figprofiles}). 

When fitting the model to the experimental data, we assumed that the number of excitons generated per second $N_p$ increases linearly with the excitation power.
At lowest power $P_0=70$~nW we have chosen $N_p^{0}=3\times 10^8$~s$^{-1}$. The value of  $N_p^0$ is the first fitting parameter (actually, this value depends on the choice of $u_0$, the true fitting parameter is $N_p^0u_0$). The two other fitting parameters are the electron spin relaxation time $\tau_e$ (or, equivalently, IX spin lifetime $\tau_s$) and the exciton diffusion constant $D$. 
The recombination time $\tau_r$  is determined for each set of data according to Eq. \ref{taur}. The resulting values are shown in Fig.~\ref{figP}~(a).

The comparison between the data and  the calculated PL intensity (assumed to be proportional to the density of bright excitons) is shown in Fig.~\ref{figprofiles} for three different temperatures and powers. 
One can see in Fig.~\ref{figP}~(c),(d) that both $\tau_s$ and $D$ obtained within the fitting procedure appear to be temperature and power dependent.
In the next Section we discuss this dependence and the underlying physical mechanisms.
\section{Discussion}
\label{discussion}
\begin{figure}
\centering{}\includegraphics[scale=0.45]{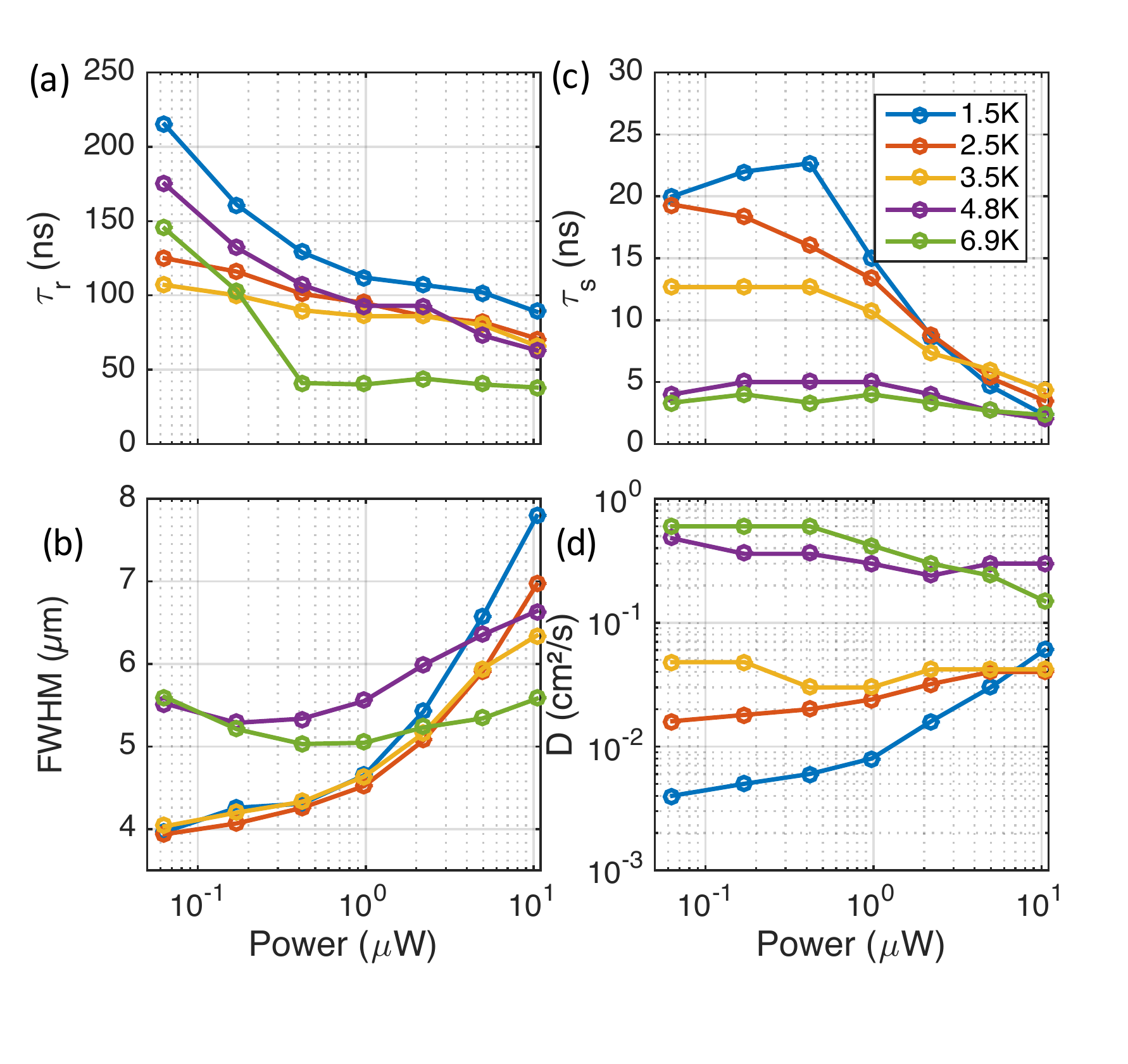}\protect\caption{ (a) IX radiative lifetime $\tau_r$ used in the modeling, see Section \ref{setup} and Eq. \ref{taur}; (b) IX cloud width at half maximum of intensity profile  as a function of excitation power for several temperatures.
IX spin lifetime $\tau_s$ (c) and exciton diffusion coefficient (d) are  extracted from the drift-diffusion modeling of the PL spatial profiles, as those shown in Fig. \ref{figprofiles}.} 
\label{figP}
\end{figure}
%
 %
%
\begin{figure}
\centering{}\includegraphics[scale=0.4]{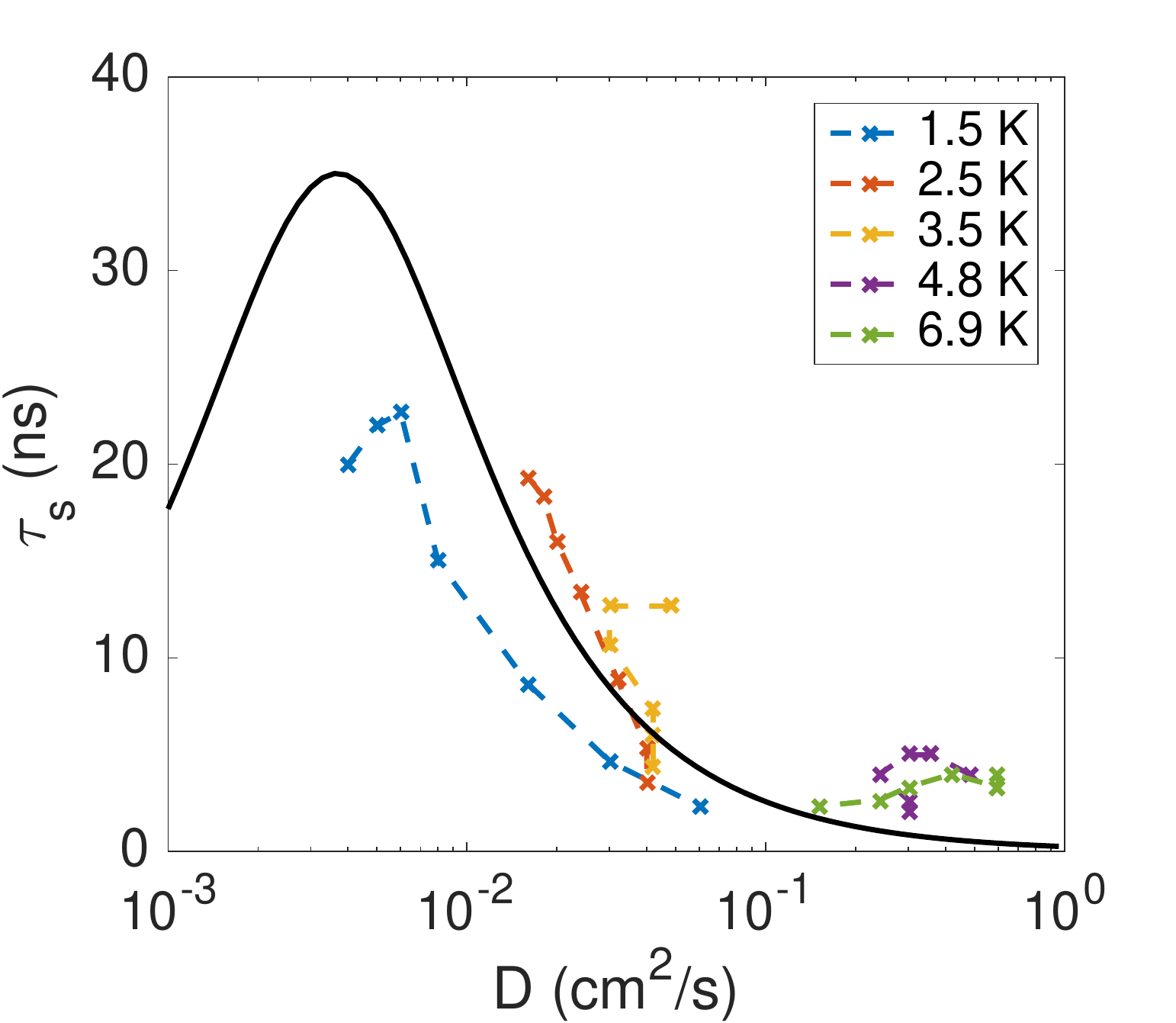}\protect\caption{ IX spin lifetime as a function of the diffusion constant as extracted from the drift-diffusion modeling. Black line is the theoretical estimation based on Eq. \ref{eq:dp_hf}.}
\label{fig:dphf}
\end{figure}
To better understand the spin dynamics of IXs, it is instructive to compare their spin lifetime to the diffusion constant, that determines the transport dynamics of the IX cloud. Both the spin and the transport show three distinct temperature regimes, as can be seen in Fig.~\ref{figP}.

(i) At  high temperature, $T \ge 4.8$~K, the spin lifetime is rather short ($\tau_s<5$~ns), and almost does not change with excitation power. At the same time, $D$ also remains approximately constant, and the size of the cloud exhibits a weak, non-monotonous behavior with power. The fact that $D$ is independent of excitation power is typical for a thermally activated cloud \cite{Rapaport2005}. The non-monotonous cloud size is due to the interplay between diffusion and drift of the excitons. It can be shown analytically that in the case of diffusion-dominated transport, the FWHM of the cloud does not depend on $N_p$, the number of particles injected in the system per unit time, but only on $(D\tau_r)^{1/2}$. By contrast, in the case of drift-dominated transport, the FWHM of the cloud increases as $\left({D N_p}/{k_B T}\right )^{1/4}\tau_r^{1/2}$. The non-monotonous behavior of the cloud size when excitation power increases could thus be the signature of transition between these two regimes \cite{Voros2005,Rapaport2006,Rapaport2007}.
%
%
%

(ii) A very different behavior is seen in the lower temperature regime, 1.5~K $<T<4.8$~K. The spin lifetime sharply rises as the temperature decreases, reaching an almost a $5$-fold increase at $2.5$~K. Spin lifetime also exhibits a strong power dependence, with a clear threshold power of $\simeq 0.5$~$\mu$W, above which it strongly decreases. This trend of $\tau_s$ is clearly anti-correlated with both the size of the cloud and the diffusion constant obtained from the modeling.
Indeed, the diffusion constant $D$ sharply decreases with decreasing temperature in this regime. 
As for the power dependence, $D$ is approximately constant with  increasing power up to the same threshold of $\simeq 0.5$~$\mu$W, and increases only above this threshold. The IX cloud does not show any expansion up to a similar excitation power, and then experience a strong expansion. Such anti-correlation of $\tau_s$ and $D$ is well explained within the DP mechanism, where the IX spin relaxation time $\tau_s$ reads: 
\begin{equation}
\tau_{s, SO}^{-1}=  \Omega_{SO}^2 \tau_p,
\end{equation}
where $\tau_p$ is the momentum scattering time,
$\Omega_{SO}= 2 \beta k $  is the electron spin precession frequency in the presence of an effective magnetic field due to spin-orbit interaction,
$\beta$ is the spin-orbit interaction constant, $k=\sqrt{2m_{exc}k_BT/\hbar^2}m_e/m_{exc}$ is the thermal wave vector of an electron within exciton, $m_e$($m_{exc}$) is the electron (exciton) mass. Because $\tau_p = D m_{exc} /k_B T$, an increase of the diffusion constant leads to a decrease of the spin relaxation time, as has been reported by Leonard {\it et al.}\cite{Leonard2009}. Note however, that the persistent localization of the IX cloud below $T=4.8$ K over a very large range of the excitation powers is consistent with previous report of a transition into a spatially localized dark liquid of IX \citep{Stern2014,Cohen2016}, which was interpreted as a condensation into a dark quantum liquid of dark spin states \cite{Shilo2013,Cohen2016,Mazuz-Harpaz2017,Alloing2013a}. In particular, Cohen \textit{et al.} observed a similar power threshold, over which there is a strong increase in IX mobility and the IX cloud expansion \cite{Cohen2016,Mazuz-Harpaz2017}.

%

(iii) Focusing on the lowest measured temperature of $1.5$~K (which also corresponds to the temperature where the transition to the liquid phase in Ref. \onlinecite{Cohen2016} was found to be fully established), a very distinct behavior of $\tau_s$ with power is observed.  
In this regime a clear increase of $\tau_s$ when the power increases is seen, together with a slow increase in $D$. This correlation between $D$ and $\tau_s$ continues up to $\simeq 0.5$ $\mu$W, after which a strong increase of $D$ is observed together with a strong decrease of $\tau_s$, where the previously discussed anti-correlation between $\tau_s$ and $D$ is established again. This low power regime of correlation can not be understood within DP mechanism and suggests that the hyperfine interaction with the nuclear spins of the GaAs lattice is involved in the spin relaxation process\cite{Dzhioev2002,Chen2007,KavokinSST}. Indeed, the fluctuating Overhauser field $B_N$ created by the nuclear spins induces a spin relaxation given by a formula very similar to the one corresponding to the DP mechanism:
\begin{equation}
\tau_{s, N}^{-1}= \frac{1}{3}\Omega_{N}^2 \tau_c, 
\end{equation}
where $\Omega_{N} = g \mu_B B_N$ is the electron spin precession frequency due to the fluctuating Overhauser field, $\tau_c$ is the correlation time of this field and $g$ is the electron $g$-factor.
The correlation time $\tau_c$ is related to the exciton diffusion coefficient and can be estimated as
the time over which an exciton moves by a distance corresponding to its own radius $a_{exc}$:
$\tau_c = a_{exc}^2/ D$. 
In contrast to the spin relaxation induced by spin-orbit interaction,  hyperfine induced spin-relaxation is enhanced when exciton diffusion slows down. Therefore, this relaxation mechanism could explain the initial increase of $\tau_s$ with power, as observed at $T=1.5$ K. 

This is the first time, to our knowledge, that such spin relaxation mechanism is observed in an excitonic system. It stems from the very low mobility regime achieved for the IX cloud, allowing for the hyperfine interaction induced dephasing to become dominant. As the power increases, there is a  transition from the hyperfine interaction to the DP dephasing mechanism, corresponding to the sharp increase in the IX diffusivity. While here we did not carefully map the state of the IX fluid as a function of power and temperature as was done in Ref.~\onlinecite{Cohen2016}, the temperature and power regimes where strong IX fluid localization and hyperfine-induced spin relaxation is observed in this work, seems to overlap with the regime where the IX liquid formation was observed. We thus can speculate that the reason for this strong localization of the IX cloud could be the formation of a dark IX liquid, similar to the previous reports\cite{Shilo2013,Stern2014,Cohen2016,Mazuz-Harpaz2017,Alloing2013a}. If this is indeed the case, it shows that measuring spin polarization can be a complementary, sensitive tool for mapping the collective states of IX fluids.

In order to get a unified picture of the observed phenomena, we write the total spin relaxation time, due to both hyperfine and spin-orbit relaxations, as:
\begin{equation}
\tau_s^{-1}= \tau_{s,N}^{-1}+ \tau_{s,SO}^{-1},
\label{eq:dp_hf}
\end{equation}
where $\tau_s$ clearly has a non-monotonous dependence on the diffusion constant. Fig.~\ref{fig:dphf} shows $\tau_s(D)$, as calculated from Eq.~\ref{eq:dp_hf} (solid line)
or extracted from measurements fitted by the drift-diffusion model of Eqs. ~\ref{eq:npm},\ref{eq:npm2}  (symbols). 
We assume $B_N= b_N/ \sqrt{N_N}$, where
$b_N=5.3$~T is the nuclear field at saturation \cite{OpticalOrientation} and $N_N \simeq 10^5$ is the number of nuclei within the exciton volume  $V_{exc}= \pi a_{exc}^2 L$, $L$ is the quantum well width. The $g$-factor of the $L=12$~nm GaAs quantum well is $g \simeq 0.35$ \cite{Dyakonov2008a} and the spin-orbit constant $\beta=25$~meV$\cdot\AA$ \cite{Leonard2009}. The theoretical curve $\tau_s(D)$ has a non-monotonous dependence and reaches a maximum $\tau_s^{max}= \sqrt{3/8}v_{th}/( \Omega_{SO} \Omega_{N} a_{exc})$ 
at $D^{max}=  (\Omega_N/\Omega_{SO}) a_{exc} v_{th}/\sqrt{6}$, where
$v_{th}=\sqrt{2k_BT/m_{exc}}$ is the thermal velocity  of excitons. Remarkably, this curve does not depend on temperature, because $\Omega_{SO}$ and $v_{th}$ have the same temperature dependence. Numerically one finds $\tau_s^{max} \simeq 35$~ns
and $D^{max} \simeq 3.7\times 10^{-3}$ cm$^{2}$/s, 
close to the experimental observation. The theoretical curve being universal,
it can be directly compared to the experimental data extracted from the numerical modeling at any temperature. Since there are no fitting parameters in the theory, the agreement can be considered as satisfactory.

\section{Conclusions}
\label{conclusions}

To conclude,
our analysis suggests that IX spin lifetime is limited not only by the
 spin-orbit interaction via DP mechanism, but also by the hyperfine interaction with the spins of the lattice nuclei.
%
The crossover between these two different regimes corresponds to a transition from  a high mobility state of the fluid to a very low mobility state, and may be correlated with the power and temperature regime where a condensation of a dipolar IX fluid was previously reported. 
These results suggest that the IX spin polarization could be a sensitive supplementary probe for the onset of the collective states in  dipolar fluids of excitons. 
Further work, both theoretical and experimental, is required to confirm the relation between spin dynamics and collective properties of IX liquids. 
In particular, it would be instructive to combine the constant energy scheme ($E_{DX}-E_{IX}=const$) used in Ref. \onlinecite{Cohen2016} with resonant spin polarized excitation used in this work to achieve better control over bright and dark exciton densities. Theoretically, the simplified model that we used must be extended to the case of non-thermal distribution of the  IX population between bright and dark states. The assumption of the infinitely fast hole spin relaxation must be lifted in this case. 
The comparison between such model and the full set of data obtained within the constant energy scheme will settle definitively the origin of exciton localization, and the onset of the hyperfine-induced spin relaxation. 

\section*{Acknowledgments}
We would like to acknowledge financial support from the German Deutsche Forschungsgemeinschaft (Grant No. SA-598/9), from the German Israeli Foundation (Grant No. GIF I-1277-303.10/2014), from the Israeli Science Foundation (Grant No. 1319/12) and French National Research Agency (Grant OBELIX, No. ANR-15-CE30-0020-02). The work at Princeton University was funded by the Gordon and Betty Moore Foundation through the Emergent Phenomena in Quantum Systems initiative Grant No. GBMF4420, and by the National Science Foundation Materials Research Science and Engineering Center Grant No. DMR-1420541. 


\section*{Appendix}
\subsection*{Dependence on excitation energy}
In order to determine the resonant energy levels of the DQW structure, a  photo-luminescence excitation (PLE) experiment was conducted in the following way: a tunable CW Titanium-Sapphire laser was scanned through a wavelength range covering the expected optical direct exciton resonances, while monitoring the (e1:hh1) 1S IX PL energy, intensity  and polarization. 
\begin{figure}
\centering{}\includegraphics[scale=0.35]{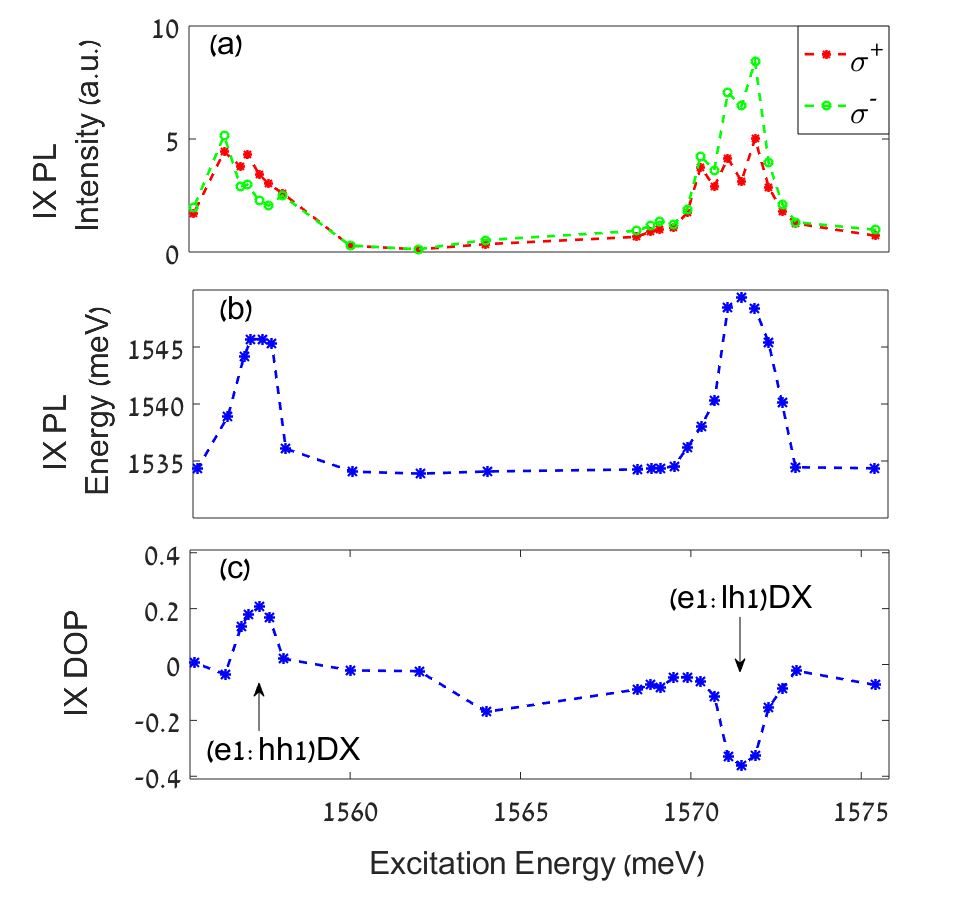}\protect\caption{Polarization-resolved PLE of IX. (a) The integrated PL intensity,
monitored at the IX line for both polarizations. (b) The
PL energy, and
(c) The degree of circular Polarization (DOP). A
negative DOP means that the PL of the IX has the opposite circular polarization with respect to the circularly polarized exciting
laser. The laser power here is 1.3 $\mu$W and is circularly polarized with $\sigma^+$. The energies of (e1:hh1) DX and (e1:lh1) DX are pointed in (c) as a guide to the eye. } \label{fig:PLE}
\end{figure}

Fig.~\ref{fig:PLE} shows the results of a polarization-resolved PLE experiment.
Fig.~\ref{fig:PLE}(a) and (b) show the integrated intensity and the peak energy of IX as a function of the excitation energy of the laser. There are two distinct peaks, separated by about 14~meV in energy. We attribute the high energy peak to the (e1:lh1) 1S DX absorption line and the lower one to the (e1:hh1) 1S DX. A large  blue shift of the IX line is observed when the excitation is at resonance with one of the DX absorption lines. This is not surprising as it is expected that the DX resonances would have a large absorption cross-section and therefore an efficient conversion of photons to IX population, by carrier tunneling to adjacent quantum wells.

Fig.~\ref{fig:PLE}(c) presents the DOP of IX as a function of the excitation energy. It can be seen that the DOP of the IX is maximal for laser excitation resonant with the DX transitions. While for excitation at the (e1:hh1) DX transition, IX are polarized in parallel to the exciting laser (positive DOP), an opposite polarization appears when the laser is resonant with the (e1:lh1) DX transition (negative DOP). This could be explained by the following mechanism: a resonant excitation
with $\sigma^{+}$ polarized light at the (e1:lh1) line creates the
following exciton state: $ \ket{\frac{1}{2},\frac{1}{2}}$. The 
light-hole in the exciton rapidly relaxes to the heavy-hole sub-band where it could
be either at $\ket{\frac{1}{2},\frac{3}{2}}$ state, which is dark or at $\ket{\frac{1}{2},-\frac{3}{2}}$ state, which is bright
and has $S_z=-1$, . Thus an initial population
of IX with polarization opposite to the laser is created. From
here on, the IX population follows the spin relaxation mechanism. Such "opposite" initial polarization was observed
in single GaAs QW by time resolved experiments but not to this extent
\cite{Dareys1993}. Another interesting feature is that the density
and the DOP following a (e1:lh1) DX excitation seem to be larger than at
a (e1:hh1) DX excitation. The increase in density is in contrast with the known oscillator strength ratios in bulk GaAs, which is three
times larger for the (e1:hh1) transition than the (e1:lh1) transition
\cite{Yu2010}.
\bibliography{final_paper_url_20170516.bib}
\end{document}